\documentclass[twocolumn,showpacs,preprintnumbers,amsmath,amssymb]{revtex4}

\usepackage{graphicx}
\usepackage{dcolumn}
\usepackage{bm}

\begin{document}

\preprint{APS/123-QED}

\title{Conditional generation of path-entangled optical NOON states}

\author{Anne E. B. Nielsen and Klaus M\o lmer}
\affiliation{Lundbeck Foundation Theoretical Center for Quantum
System Research, Department of Physics and Astronomy, University of
Aarhus, DK-8000 \AA rhus C, Denmark}

\date{\today}

\begin{abstract}
We propose a measurement protocol to generate path-entangled NOON
states conditionally from two pulsed type II optical parametric
oscillators. We calculate the fidelity of the produced states and
the success probability of the protocol. The trigger detectors are
assumed to have finite dead time, and for short pulse trigger fields
they are modeled as on/off detectors with finite efficiency.
Continuous-wave operation of the parametric oscillators is also
considered.
\end{abstract}


\maketitle

\section{Introduction}

Nonclassical states of light have many applications, and a number of
different protocols exist for the generation of various classes of
states. The two-mode maximally entangled $N$-photon states
\begin{equation}\label{NOON}
|\textrm{NOON}\rangle=\frac{1}{\sqrt{2}}\left(|N,0\rangle+
e^{i\phi}|0,N\rangle\right),
\end{equation}
the so-called NOON states, are particularly interesting because a
single-photon phase shift of $\chi$ induced in one of the two
components changes the relative phase of the two terms by $N\chi$.
This special property of NOON states may be utilized to enhance
spatial resolution in (quantum) microscopy and lithography
\cite{boto}, and in interferometry it has been shown that a certain
measurement strategy, using NOON states, leads to a phase estimation
error scaling as $L^{-1/4}N_T^{-3/4}$ if the phase to be estimated
is known to lie within an interval from $-\pi/L$ to $\pi/L$, where
$N_T$ is the total number of photons used in the measurements
\cite{pezze}. This is better than the classical shot noise limit of
$N_T^{-1/2}$, and NOON states are thus useful to perform accurate
measurements and may be a valuable field resource in sensors. NOON
states are also a source of entanglement with applications in
quantum information protocols and in fundamental studies such as
tests of Bell´s inequality \cite{dangelo}.

It is thus of great interest to be able to produce NOON states, and
various NOON state generation schemes have been suggested
theoretically \cite{kok,fiurasek,hofmann,walther,eisenberg,vanmeter}
and studied in experiments
\cite{walther,mitchell,eisenberg,ourjoumtsev1001}. The $N=1$ and
$N=2$ NOON states may be generated by combining either a single
photon and a vacuum state or two single-photon states on a $50:50$
beam splitter, but this simple approach is not directly extendable
to $N>2$, and we shall thus mainly be concerned with generation of
$N=3$ NOON states in the present paper, even though the suggested
protocol is, in principle, applicable for all $N$. Mitchell,
Lundeen, and Steinberg have generated NOON states with $N=3$ from a
pair of down converted photons and a local oscillator photon using
certain polarization transforming components and post-selection
\cite{mitchell}. In this experiment, however, the successful
generation of the NOON state is witnessed by a destructive detection
of the state. In the present paper we propose and analyze in detail
a nondestructive generation protocol, which conditions the
successful generation of the $N$-photon NOON state on the
registration of $N$ photo detection events in other field modes, and
which uses as resource only linear optics and the output from two
optical parametric oscillators (OPOs). The protocol does not rely on
efficient photo detection. The analysis is carried out in terms of
Wigner function formalism, and effects of finite detector efficiency
and finite detector dead time are considered.

Conditional generation of nonclassical states occupying a single
mode has been investigated both experimentally and theoretically
\cite{dakna,uren,ourjoumtsevcat,moelmer,ourjoumtsev2,neergaard,wakui,nielsen1,nielsen2}.
With the correlated output from a single nondegenerate OPO it is,
for instance, possible to generate $n$-photon Fock states of light
in the signal beam conditioned on $n$ photo detections in the idler
(trigger) beam \cite{ourjoumtsev2,nielsen1,nielsen2}, and in
principle the entanglement of a highly squeezed two-mode field from
an OPO makes it possible to prepare any state in the signal beam
that can be either measured as an eigenstate of a suitable
observable of the idler beam or produced as the final state of a
generalized measurement. The basic idea of the protocol proposed in
the present paper is to mix the output from two OPOs and employ the
entanglement to prepare a two-mode state in two of the output beams
by detection of the desired output state in the remaining beams.

In Sec.\ II we explain the NOON state generation protocol in detail.
In Sec.\ III we analyze the performance of the protocol
quantitatively for pulsed OPO sources. We provide the fidelity of
the generated states and the success probability. In Sec.\ IV we
consider production of NOON states from continuous-wave OPO sources,
and Sec.\ V concludes the paper.

\section{Experimental setup for NOON state generation}

\begin{figure}
\begin{center}
\includegraphics*[viewport=111 93 320 195,width=0.80\columnwidth]{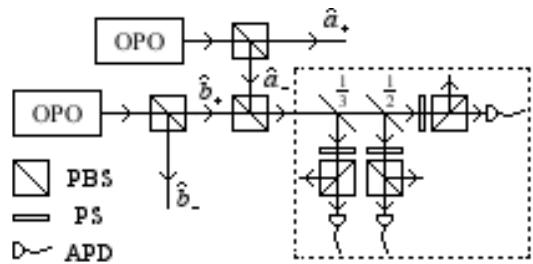}
\end{center}
\caption{Experimental setup for NOON state generation. OPO, optical
parametric oscillator; PBS, polarizing beam splitter; PS, phase
shifter; and APD, avalanche photo diode. The part of the setup
enclosed in the dashed box performs the NOON state measurement, and
here it is shown for $N=3$. Note that the polarizing beam splitters
inside the box are oriented at $45^{\circ}$. The numbers denote beam
splitter reflectivities of 1/3 and 1/2, and the three phase shifters
transform $\hat{b}_+$ into $-\hat{b}_+e^{2\pi in/3}$, where
$n=1,2,3$, respectively. See text for details.} \label{setup}
\end{figure}

The experimental setup is illustrated in Fig.\ \ref{setup}. Two
pulses of two-mode squeezed states are generated by two identical
OPOs via type II parametric down conversion. The field mode
operators of the modes generated by the first OPO are denoted
$\hat{a}_+$ and $\hat{a}_-$, respectively, while the field mode
operators of the modes generated by the second OPO are denoted
$\hat{b}_+$ and $\hat{b}_-$, respectively. For definiteness, we
assume that the plus modes are vertically polarized and that the
minus modes are horizontally polarized. The modes are separated
spatially by the first two polarizing beam splitters, and the third
polarizing beam splitter combines the $\hat{a}_-$ and $\hat{b}_+$
modes, which are subsequently subjected to the NOON state
measurement proposed in \cite{sun} and illustrated for $N=3$ in
Fig.\ \ref{setup}. The idea behind this measurement is to apply the
highly nonlinear operator
$\hat{A}_N=\hat{a}_-^N-(\hat{b}_+e^{i\theta})^N$ to the state. The
result is only nonzero if either the $\hat{a}_-$ mode or the
$\hat{b}_+$ mode contains at least $N$ photons. On the other hand,
if the squeezing is sufficiently small, it is unlikely to have more
than a total of $N$ photons in the two trigger modes, and by
conditioning on the successful application of $\hat{A}_N$, we select
the pulses of the system where $N$ photon pairs are generated in one
OPO and zero photon pairs in the other. It is equally probable that
the photons originate from the first OPO or from the second OPO,
and, as we shall see in detail below, the result is that a NOON
state is generated conditionally in the output modes $\hat{a}_+$ and
$\hat{b}_-$.

As stated in \cite{sun}, $\hat{A}_N$ can be rewritten as a simple
product of single photon annihilation operators
\begin{equation}\label{operator}
\hat{a}_-^N-\left(\hat{b}_+e^{i\theta}\right)^N=\prod_{n=1}^N
\left(\hat{a}_--\left(\hat{b}_+e^{i\theta}\right)e^{i2\pi
n/N}\right),
\end{equation}
and it is thus possible to implement $\hat{A}_N$ by means of beam
splitters and photo detectors. We first consider odd values of $N$.
Beam splitters are used to divide the input into $N$ distinct
spatial modes labeled by $n=1,\ldots,N$. The beam splitter
reflectivities are chosen to obtain the same expectation value of
the intensity in each of the modes. The vertically polarized modes
are then phase shifted by the factor $e^{i2\pi n/N+i\pi}$ relative
to the horizontally polarized modes, i.e.,
$\hat{b}_+\rightarrow-\hat{b}_+e^{i2\pi n/N}$, and finally
polarizing beam splitters with principal planes oriented at
$45^{\circ}$ relative to the horizontal polarization transform
$\hat{a}_-$ and $-\hat{b}_+e^{i2\pi n/N}$ into
$(\hat{a}_--\hat{b}_+e^{i2\pi n/N})/\sqrt{2}$ (the transmitted mode)
and $(\hat{a}_-+\hat{b}_+e^{i2\pi n/N})/\sqrt{2}$ (the reflected
mode) \cite{hariharan}. The annihilation of a photon in each of the
modes transmitted by the beam splitters witnesses the overall
application of the operator $\hat{A}_N$. If one observes both
reflected and transmitted modes simultaneously, one conditions on
detection events in all the transmitted modes and no detection
events in all the reflected modes. If detection events are instead
observed in all the reflected modes and in none of the transmitted
modes, an operator of the form \eqref{operator} is also obtained,
but $\theta$ is effectively transformed into $\theta+\pi$ due to the
phase shift at the polarizing beam splitter, and the value of $\phi$
of the generated NOON states is changed by $N\pi$ (see below). The
success probability is thus increased by a factor of two if both
outcomes are accepted.

For even values of $N$ a similar measurement scheme is applicable,
but it is sufficient to divide the field into $N/2$ spatial modes
initially, and in this case the NOON state generation is conditioned
on detection events in both transmitted and reflected modes (see
\cite{sun}).

\section{Performance of the protocol}

After this presentation of the basic idea and the physical setup we
now consider the actual outcome of the detection process. For short
pulse OPO output the dead time of the photo detectors may typically
be longer than the pulse duration, and we shall thus assume that it
is impossible to obtain more than a single detection event per
detector per pulse, i.e., if the detector efficiency is unity, the
detectors are only able to distinguish between vacuum and states
different from the vacuum state. Such detectors are denoted on/off
detectors, and they are discussed in detail in Ref.\ \cite{rohde}.
The finite dead time of the detectors is not severe to the
measurement procedure described in \cite{sun} because the on/off
detector model and the conventional photo detector model,
represented by the annihilation operator, lead to identical signal
states if the total number of photons in the idler modes is
guaranteed to be less than or equal to the number of conditioning
detection events, i.e., $N$.

We analyze the performance of the setup using Gaussian Wigner
function formalism \cite{moelmer,nielsen1,nielsen2}, which is
applicable because the squeezed states generated by the OPOs and the
vacuum states coupled into the system via the beam splitters are all
Gaussian. In general, the Wigner function of an $n$-mode Gaussian
state with zero mean field amplitude takes the form
\begin{equation}\label{gaus}
W_V(x_1,p_1,\ldots,x_n,p_n)=\frac{1}{\pi^n\sqrt{det(V)}}e^{-y^TV^{-1}y},
\end{equation}
where $y\equiv(x_1,p_1,\ldots,x_n,p_n)^T$ and $V$ is the
$2n\times2n$ covariance matrix. If $\hat{c}_i$ denotes the field
mode annihilation operator of mode $i$, the elements of $V$ are
given in terms of the real and imaginary parts of the expectation
values $\langle\hat{c}_i^\dag\hat{c}_j\rangle$ and
$\langle\hat{c}_i\hat{c}_j\rangle$. Note that for a multi-mode
Gaussian state we are free to include only the modes of interest in
\eqref{gaus} because the partial trace operation over unobserved
modes is equivalent to integration over the corresponding quadrature
variables. A unit efficiency `on' detection in mode $i$ projects
mode $i$ on the subspace of Hilbert space that is orthogonal to the
vacuum state, i.e., the Wigner function is multiplied by $(1-2\pi
W_0(x_i,p_i))$, where $W_0(x,p)=\exp(-x^2-p^2)/\pi$ is the Wigner
function of the vacuum state, the variables $x_i$ and $p_i$ are
integrated out, and the state is renormalized. Since the Gaussian
nature of a state is preserved under linear transformations, and
since a detector with single-photon efficiency $\eta$ is equivalent
to a beam splitter with transmission $\eta$ followed by a unit
efficiency detector \cite{rohde}, effects of non-unit detector
efficiency are easily included in the covariance matrix.

To calculate $\langle\hat{c}_i^\dag\hat{c}_j\rangle$ and
$\langle\hat{c}_i\hat{c}_j\rangle$ explicitly we note that the state
generated by the OPOs is \cite{ekert}
\begin{equation}\label{psii}
|\psi_\textrm{i}\rangle=(1-r^2)\sum_{n=0}^\infty\sum_{m=0}^\infty
r^{n+m}|n,n,m,m\rangle,
\end{equation}
where $r$ is the squeezing parameter and the modes are listed in the
order: $\hat{a}_+$, $\hat{a}_-$, $\hat{b}_+$, $\hat{b}_-$. We assume
that $N$ is odd and consider the transmitted trigger modes (which we
number from $1$ to $N$), the $\hat{a}_+$ mode (mode $N+1$), and the
$\hat{b}_-$ mode (mode $N+2$). By expressing the field operators of
the trigger modes (those observed by the unit efficiency detectors)
in terms of $\hat{a}_-$, $\hat{b}_+$, and field operators
representing vacuum states we find
\begin{eqnarray}
&\langle\hat{c}_j^\dag\hat{c}_k\rangle&=
\langle\psi_i|\sqrt{\frac{\eta}{2N}}(\hat{a}_-^\dag-e^{-2\pi
ij/N}\hat{b}_+^\dag e^{-i\theta})\nonumber\\
&&\hspace{1.5cm}\sqrt{\frac{\eta}{2N}}(\hat{a}_--e^{2\pi
ik/N}\hat{b}_+e^{i\theta})|\psi_i\rangle\nonumber\\
&&=\left(1+e^{2\pi i(k-j)/N}\right)\frac{\lambda}{2N},
\end{eqnarray}
where $j\in\{1,2,\ldots,N\}$, $k\in\{1,2,\ldots,N\}$,
$\lambda\equiv\eta r^2/(1-r^2)$, and we allow of a constant phase
shift $\theta$ of $\hat{b}_+$ relative to $\hat{a}_-$. Furthermore
\begin{eqnarray}
&\langle\hat{c}_{N+1}^\dag\hat{c}_{N+1}\rangle&=
\langle\psi_i|\hat{a}_+^\dag\hat{a}_+|\psi_i\rangle=r^2/(1-r^2),\\
&\langle\hat{c}_{N+2}^\dag\hat{c}_{N+2}\rangle&=
\langle\psi_i|\hat{b}_-^\dag\hat{b}_-|\psi_i\rangle=r^2/(1-r^2),\\
&\langle\hat{c}_k\hat{c}_{N+1}\rangle&=
\langle\psi_i|\sqrt{\frac{\eta}{2N}}(\hat{a}_--e^{2\pi
ik/N+i\theta}\hat{b}_+)\hat{a}_+|\psi_i\rangle\nonumber\\
&&=\sqrt{\frac{\eta}{2N}}\frac{r}{1-r^2},\\
&\langle\hat{c}_k\hat{c}_{N+2}\rangle&=
\langle\psi_i|\sqrt{\frac{\eta}{2N}}(\hat{a}_--e^{2\pi
ik/N+i\theta}\hat{b}_+)\hat{b}_-|\psi_i\rangle\nonumber\\
&&=-\sqrt{\frac{\eta}{2N}}\frac{r}{1-r^2}e^{2\pi ik/N+i\theta},
\end{eqnarray}
and
\begin{multline}
\langle\hat{c}_j\hat{c}_k\rangle=
\langle\hat{c}_{N+1}\hat{c}_{N+1}\rangle=
\langle\hat{c}_{N+2}\hat{c}_{N+2}\rangle=
\langle\hat{c}_{N+1}\hat{c}_{N+2}\rangle=\\
\langle\hat{c}_{N+1}^\dag\hat{c}_{N+2}\rangle=
\langle\hat{c}_k^\dag\hat{c}_{N+1}\rangle=
\langle\hat{c}_k^\dag\hat{c}_{N+2}\rangle=0.
\end{multline}
For even values of $N$ the factors $\sqrt{\eta/(2N)}$ are replaced
by $\sqrt{\eta/N}$. Note that loss in the signal beam may be taken
into account by performing the transformations
$\hat{a}_+\rightarrow\sqrt{\eta_s}\hat{a}_+$ and
$\hat{b}_-\rightarrow\sqrt{\eta_s}\hat{b}_-$ in the above
expressions, where $1-\eta_s$ is the loss.

\begin{figure}
\begin{center}
\includegraphics*[viewport=19 7 392 298,width=0.85\columnwidth]{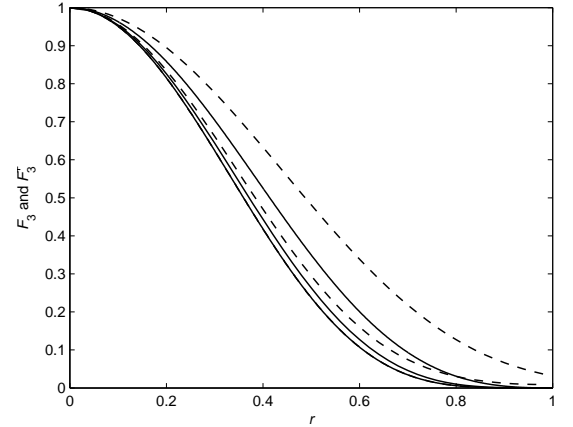}
\end{center}
\caption{NOON state fidelity $F_3$ (solid lines) and $F_3^+$ (dashed
lines) as a function of squeezing parameter $r$ for $\eta=1$ (upper
lines), $\eta=0.25$ (middle lines), and $\eta\rightarrow0$ (lower
lines). Note that in the latter case $F_3=F_3^+$.} \label{rF}
\end{figure}

\begin{figure}
\begin{center}
\includegraphics*[viewport=8 7 392 300,width=0.85\columnwidth]{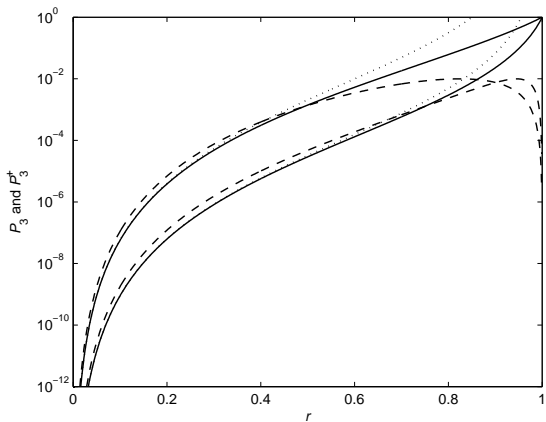}
\end{center}
\caption{Success probability $P_3$ (solid lines) and $P_3^+$ (dashed
lines) as a function of squeezing for $\eta=1$ (upper lines) and
$\eta=0.25$ (lower lines). The dotted lines represent the
approximate expression \eqref{Pap}.} \label{rP}
\end{figure}

The NOON state fidelity $F_N$ of the signal state conditioned on $N$
photo detection events in the transmitted trigger modes is
\begin{multline}\label{FNi}
F_N=\frac{4\pi^2}{P_N}\int W_\textrm{NOON}(x_{N+1},p_{N+1},x_{N+2},p_{N+2})\\
\left(\prod_{i=1}^N\left(1-2\pi W_0(x_i,p_i)\right)\right)\\
W_V(x_1,p_1,\ldots,x_{N+2},p_{N+2})
\left(\prod_{i=1}^{N+2}dx_idp_i\right),
\end{multline}
where $W_\textrm{NOON}$ is the Wigner function of the NOON state
\eqref{NOON}, and
\begin{multline}\label{PNi}
P_N=\int \left(\prod_{i=1}^N\left(1-2\pi W_0(x_i,p_i)\right)\right)\\
W_V(x_1,p_1,\ldots,x_{N+2},p_{N+2})
\left(\prod_{i=1}^{N+2}dx_idp_i\right),
\end{multline}
is the success probability, i.e., the probability to obtain the
conditioning detection events and produce the NOON state in a given
pulse of the OPO system. We expand the product
\begin{equation}
\prod_{i=1}^N\left(1-2\pi W_0(x_i,p_i)\right)=
\sum_d\prod_{i=1}^N(-2\pi W_0(x_i,p_i))^{d_i},
\end{equation}
where the sum is over all $d\equiv(d_1,d_2,\ldots,d_N)$ with
$d_i\in\{0,1\}$, and define the diagonal matrix
$J_d=\mathrm{diag}(d_1,d_1,d_2,d_2,\ldots,d_N,d_N)$ and the $n\times
n$ identity matrix $I_n$. Furthermore, we divide the covariance
matrix into four parts
\begin{equation}
V=\left[\begin{array}{cc}V_{tt}&V_{ts}\\V_{ts}^T&V_{ss}\end{array}\right],
\end{equation}
where $V_{tt}$ is the $2N\times2N$ covariance matrix of the trigger
modes, $V_{ss}$ is the $4\times4$ covariance matrix of the signal
modes, while $V_{ts}$ contains the correlations between the trigger
and the signal modes, and we define the vector
$y_s=(x_{N+1},p_{N+1},x_{N+2},p_{N+2})^T$ and the matrix
\begin{equation}
U_d=V_{ss}-V_{ts}^TJ_d(J_dV_{tt}J_d+I_{2N})^{-1}J_dV_{ts}.
\end{equation}
This allows us to write Eqs.\ \eqref{FNi} and \eqref{PNi} in the
following compact forms \cite{eisert}
\begin{multline}\label{FN}
F_N=\frac{4\pi^2}{P_N}\sum_d\frac{(-2)^{\sum_{i=1}^Nd_i}}
{\sqrt{\det(I_{2N}+J_dV_{tt})}}\\
\int W_\textrm{NOON}(y_s)W_{U_d}(y_s)dy_s,
\end{multline}
and
\begin{equation}\label{PN}
P_N=\sum_d\frac{(-2)^{\sum_{i=1}^Nd_i}}{\sqrt{\det(I_{2N}+J_dV_{tt})}}.
\end{equation}
Since $W_\textrm{NOON}$ is a product of a polynomial and a Gaussian
the integral in Eq.\ \eqref{FN} may be evaluated analytically and
for $N=3$ and $\eta=1$ we find
\begin{equation}
F_3^{\eta=1}=\frac{(1-r^2)^2(2-r^2)^2(3-2r^2)(6-5r^2)}{18(4-3r^2)},
\end{equation}
where the optimal value $\phi=N\theta+\pi+2\pi n$, $n\in Z$, is
assumed. Expressions for $P_N$ are given in table \ref{tabel} for
$N=1,2,3,$ and $4$, and $F_N$ and $P_N$ are plotted for $N=3$ in
Figs.\ \ref{rF} and \ref{rP}, respectively. We observe that high
probabilities are only found in the parameter regime, where the
fidelity is low. If, for instance, we want a NOON state fidelity of
at least $0.9$, we choose $r=0.14$, and if $\eta=0.25$, $P_3$ is of
order $10^{-8}$. With a repetition rate of order $10^6\textrm{
s}^{-1}$ (see \cite{ourjoumtsev2}) one state is produced every
second minute on average. The production rate is very dependent on
detector efficiency, and if $\eta$ is increased to unity, the rate
is increased by approximately a factor of $60$.

\begin{table}
\begin{tabular}{|c|c|c|}
\hline $N$ & $P_N$ & $P_N^+$\\
\hline 1 & $\frac{\lambda}{\lambda+1}$ & $\frac{2\lambda}{(\lambda+1)^2}$\\
\hline 2 & $\frac{\lambda^2}{(\lambda+1)^2}$ & -\\
\hline 3 & $\frac{\lambda^3(\lambda+4)}
{(\lambda+2)^2(\lambda+3)(\lambda+6)}$ &
$\frac{2\lambda^3(3\lambda+4)}{
(\lambda+1)^2(\lambda+2)^2(2\lambda+3)(5\lambda+6)}$\\
\hline 4 & $\frac{\lambda^4(\lambda^2+6\lambda+6)}
{(\lambda+1)^2(\lambda+2)^2(\lambda^2+8\lambda+8)}$ & -\\
\hline
\end{tabular}
\caption{Success probabilities calculated from Eqs.\ \eqref{PN} and
\eqref{PNp}. $\lambda\equiv\eta r^2/(1-r^2)$.}\label{tabel}
\end{table}

For odd values of $N$ we may observe both reflected and transmitted
trigger modes and condition on detection events in all the
transmitted trigger modes and no detection events in all the
reflected trigger modes, or, {\it vice versa}. In this case we also
include the reflected trigger modes in the covariance matrix, which
we now denote by $V^+$. By a similar analysis as above we obtain the
success probability
\begin{equation}\label{PNp}
P_N^+=2\sum_d\frac{2^N(-2)^{\sum_{i=1}^Nd_i}}{\sqrt{\det(I_{4N}+J^+_dV^+_{tt})}},
\end{equation}
where
$J^+_d\equiv\textrm{diag}(d_1,d_1,\ldots,d_N,d_N,1,1,\ldots,1,1)$,
while the NOON state fidelity $F_N^+$ is given by Eq.\ \eqref{FNi}
with $V$ replaced by the matrix
\begin{equation}
V-(V^+_R)^T(V^+_{RR}+I_{2N})^{-1}V^+_R,
\end{equation}
where $V^+_{RR}$ is the covariance matrix of the reflected trigger
modes and $V^+_R$ consists of the correlations between the reflected
trigger modes and the signal and transmitted trigger modes. Explicit
results for $P^+_N$ are given in table \ref{tabel} for $N=1$ and
$3$. $F_3^+$ and $P_3^+$ are compared to $F_3$ and $P_3$ in Figs.\
\ref{rF} and $\ref{rP}$, and it is observed that $F_3^+$ and $P_3^+$
are both larger than $F_3$ and $P_3$ if $r$ is not large (and
$\eta>0$). For $r\rightarrow1$, $P_3^+\rightarrow0$ because in this
limit it is very unlikely to obtain no detection events in all the
reflected or in all the transmitted trigger modes.

In the limit of very small detector efficiency a simple expression
for the NOON state fidelity for the case of $N$ trigger detectors is
easily derived without using Wigner function formalism. In general,
if the state of interest is expressed in the photon number basis,
the mathematical operation corresponding to an `on' detection is to
multiply each term by $\sqrt{1-(1-\eta)^n}$, where $n$ is the number
of photons in the mode observed by the non-unit efficiency detector,
trace out the detected mode, and renormalize. If $n\eta\ll1$ for all
contributing terms,
$\sqrt{1-(1-\eta)^n}\approx\sqrt{n\eta}\propto\sqrt{n}$, and the
on/off detector model becomes equivalent to the photo detector
model. In this case the density operator of the output state is
obtained as
\begin{eqnarray}
\rho&=&M\sum_{p=0}^\infty\sum_{q=0}^\infty\langle p|\langle q|
(\hat{a}_-^N-(\hat{b}_+e^{i\theta})^N)|\psi_i\rangle\nonumber\\
&&\hspace{2.3cm}\langle\psi_i| ((\hat{a}_-^\dag)^N-(\hat{b}_+^\dag
e^{-i\theta})^N)
|q\rangle|p\rangle\nonumber\\
&=&\frac{(1-r^2)^{N+2}}{2N!r^{2N}}\nonumber\\
&&\bigg(\sum_{n=N}^\infty\sum_{m=0}^\infty
(r^2)^{n+m}\Big(\frac{n!}{(n-N)!}|n,m\rangle\langle
n,m|\nonumber\\
&&-e^{-iN\theta}
\sqrt{\frac{n!(m+N)!}{(n-N)!m!}}|n,m\rangle\langle n-N,m+N|\Big)\nonumber\\
&&+\sum_{n=0}^\infty\sum_{m=N}^\infty(r^2)^{n+m}
\Big(\frac{m!}{(m-N)!}|n,m\rangle\langle n,m|\nonumber\\
&&-e^{iN\theta}\sqrt{\frac{(n+N)!m!}{n!(m-N)!}}|n,m\rangle\langle
n+N,m-N|\Big)\bigg),\nonumber\\
\end{eqnarray}
where $M$ is a normalization constant and the traces are over the
$\hat{a}_-$ and $\hat{b}_+$ modes. This leads to the NOON state
fidelity
\begin{equation}\label{FNOON}
F_N^{\eta\ll1}=
\langle\textrm{NOON}|\rho|\textrm{NOON}\rangle=(1-r^2)^{N+2},
\end{equation}
where again $\phi=N\theta+\pi+2\pi n$, $n\in Z$, is assumed. It is
interesting to compare this result with the fidelity $(1-r^2)^{N+1}$
obtained for production of $N$-photon states from a single two-mode
squeezed state by conditioning on $N$ detection events in the idler
beam and using detectors with very small efficiency. If a
single-photon state is produced by this method and transformed into
an $N=1$ NOON state as explained in the Introduction, the NOON state
fidelity is $F_{1,s}=(1-r^2)^2$, and the success probability is
$P_{1,s}=\lambda/(\lambda+1)$. Choosing squeezing parameters such
that $F_{1,s}=F_1$, we find that $P_1^+=(4/3)P_{1,s}$ in the high
fidelity limit. It is thus possible to achieve a higher success
probability using the scheme with two OPOs, but the price to pay is
a more technically involved setup, and NOON states with two
different values of $\phi$ are produced. For $N=2$ the present
protocol and combination of two single-photon states on a $50:50$
beam splitter, each produced conditionally from a single OPO, lead
to identical fidelities and success probabilities. Finally we note
that the photo detector model underestimates $F_N$ for $\eta>0$
because $1-(1-\eta)^n=\eta\sum_{i=0}^{n-1}(1-\eta)^i<n\eta$ for
$n=2,3,\ldots$ while $1-(1-\eta)^n=n\eta$ for $n=0,1$, i.e., the
`wrong' terms containing more than $N$ photons are given a too large
weight. This is also what we observe in Fig.\ \ref{rF}.

In the limit of small $r$ and for odd values of $N$ the success
probability is given approximately by the simple expression
\begin{multline}\label{Pap}
P_N\approx\left(\frac{\eta}{2N}\right)^N\langle\psi_i|
((\hat{a}_-^\dag)^N-(\hat{b}_+^\dag e^{-i\theta})^N)\\
(\hat{a}_-^N-(\hat{b}_+e^{i\theta})^N)|\psi_i\rangle
=\frac{2N!}{(2N)^N}\lambda^N\textrm{ (}N\textrm{ odd)}.
\end{multline}
Again $\eta/(2N)$ must be replaced by $\eta/N$ to obtain $P_N$ for
even values of $N$. The approximation to $P_3$ is shown in Fig.\
\ref{rP}.

\section{NOON states from continuous-wave OPO sources}

Our protocol is not limited to pulsed fields, and for completeness
we now briefly consider NOON state generation from continuously
driven OPOs. We assume $N=3$. For continuous-wave fields each of the
three detected trigger beams and the two signal beams are described
by time dependent field operators $\hat{c}_i(t)$. The trigger
detections take place in particular modes localized around the three
detection times $t_{c1}$, $t_{c2}$, and $t_{c3}$, and we want to
determine the NOON state fidelity of an output state occupying one
temporal mode in each signal beam. Following the general multimode
formalism in Refs.\ \cite{moelmer,nielsen2}, the five relevant modes
are specified by the mode functions $f_i(t)$, and the corresponding
five single mode operators are $\hat{c}_i=\int
f_i(t)\hat{c}_i(t)dt$. In general, we are free to choose the two
output mode functions at will, and in the present case it is natural
to choose the mode function which gives rise to the largest
three-photon state fidelity when three-photon states are generated
from a single type II continuous-wave OPO. Since we are mainly
interested in the parameter region where the squeezing is small and
the NOON state fidelity is large, we use the optimal three-photon
state mode function derived for very low beam intensity in
\cite{nielsen2}, i.e.,
$f_4(t)=f_5(t)=\sum_{k=1}^3c_k\sqrt{\gamma/2}\exp(-\gamma|t-t_{ck}|/2)$,
where the coefficients $c_k$ are functions of the intervals between
the detection times and $\gamma$ is the leakage rate of the OPO
output mirror. We furthermore assume that the trigger mode functions
are nonzero only in an infinitesimal time interval centered at the
detection time, which is valid if the trigger detections take place
on a time scale much shorter than $\gamma^{-1}$. Since we consider a
low intensity continuous beam, and since we formally assume that the
trigger detectors only register the light field in infinitesimal
time intervals around the detection times, the annihilation operator
detector model is perfectly valid in this case and detector dead
time is insignificant.

\begin{figure}
\begin{center}
\includegraphics*[viewport=19 7 392 300,width=0.85\columnwidth]{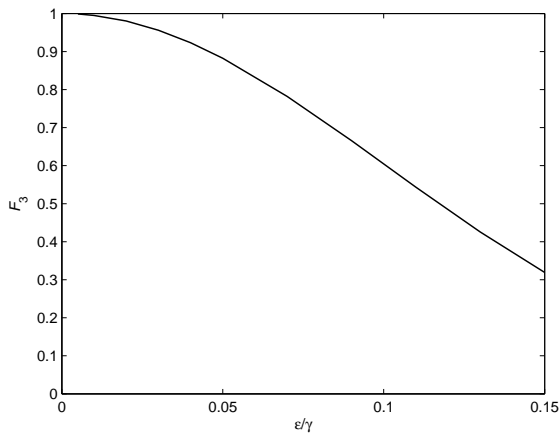}
\end{center}
\caption{NOON state fidelity as a function of $\epsilon/\gamma$ for
states generated from a pair of continuous-wave OPO sources when
conditioning on three simultaneous trigger detection events
$t_{c1}=t_{c2}=t_{c3}$.} \label{eF}
\end{figure}

\begin{figure}
\begin{center}
\includegraphics*[viewport=19 7 385 298,width=0.85\columnwidth]{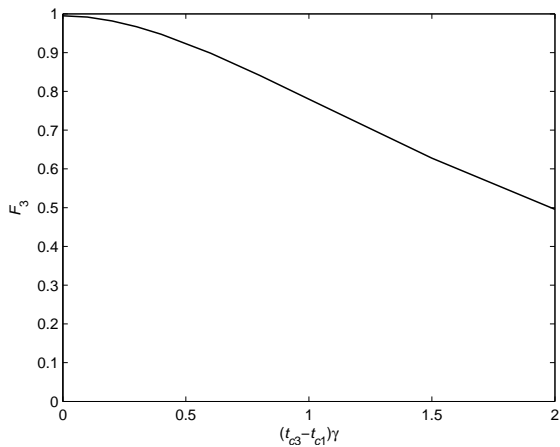}
\end{center}
\caption{Fidelity of NOON states from continuous-wave OPO sources as
a function of separation between trigger detection events
$(t_{c3}-t_{c1})\gamma$ for $N=3$, $t_{c3}-t_{c2}=t_{c2}-t_{c1}$,
and $\epsilon/\gamma=0.01$.} \label{tF}
\end{figure}

We may now proceed as above and eliminate all the irrelevant modes
from the analysis by writing down the Gaussian Wigner function of
the five interesting modes. The only difference is that this time
$\langle\hat{c}_i^\dag\hat{c}_j\rangle$ and
$\langle\hat{c}_i\hat{c}_j\rangle$ are expressed in terms of the two
time correlation functions of the OPO output. Also, the operators
applied to the Wigner function to take conditioning into account are
different because the annihilation detector model is used. The
reader is referred to Refs.\ \cite{moelmer,nielsen2} for details.

The resulting fidelity is shown as a function of $\epsilon/\gamma$
in Fig.\ \ref{eF}, where $\epsilon$ is the nonlinear gain in the
OPO, and as a function of the temporal distance between the
conditioning detection events in Fig.\ \ref{tF}. As in the pulsed
case the fidelity decreases when the degree of squeezing increases.
The fidelity also decreases when the temporal distance between the
conditioning detection events increases from zero, but it is
permissible to have a small time interval between the trigger
detection events. We note that the curves represent a lower limit to
the theoretically achievable fidelity since a better fidelity may be
obtained for another choice of output state mode functions.

\section{Conclusion}
In conclusion we have analyzed a method to generate path entangled
NOON states from the output from two optical parametric oscillators.
The method relies on the joint detection of photons in a number of
trigger beams, and we presented a theoretical analysis of the role
of detector efficiency and dead time for the fidelity of the states
obtained and the success probability of the protocol. Our specific
NOON state protocol applies for general photon numbers of the
states, but in practice it is not realistic to go beyond the case of
$N=3$, studied here. This is due to the reduction of the fidelity
due to unwanted contributions from higher number states, when the
OPO output power gets too high, combined with the severe reduction
of the probability to obtain the number of conditioning detection
events needed when the OPO output power is too low. The $N=3$ NOON
states, which can be produced at $90\%$ fidelity at the rate of one
state produced every $10-100$ seconds, seem to be at the limit of
realistic experiments of the proposed kind. Finally, we also
determined the NOON state fidelity for continuous-wave fields, where
the best NOON state occupies a pair of suitably selected temporal
mode functions, and where we find high fidelities as long as the
trigger events occur within a short time window compared to the
lifetime of the OPO cavity field.

We presented this analysis for the production of optical NOON
states, but we note that recent theoretical proposals and
experiments with four wave mixing of matter waves \cite{campbell},
engineered quadratic interactions among trapped ions \cite{cirac},
and entanglement between field and atomic degrees of freedom
\cite{monroe,harald} bring promise for similar conditional
generation of atomic and interspecies atom-field NOON states.

This work was supported by the European Integrated project SCALA.

\end{document}